\documentclass[doublecol]{epl2}
\usepackage[utf8x]{inputenc}
\usepackage{amsmath,amssymb,graphicx,url,color}

\newcommand{\avg}[1]{\langle #1\rangle}
\newcommand{\eg}{\emph{e.g.}}
\newcommand{\ie}{\emph{i.e.}}
\newcommand{\neff}{n_{\mathrm{eff}}}
\newcommand{\Nd}{\mathcal{N}}
\title{Crowd Avoidance and Diversity in Socio-Economic Systems and Recommendation}
\author{S. Gualdi\inst{1} \and M. Medo\inst{1} \and Y.-C. Zhang\inst{1,2}}
\shortauthor{S. Gualdi \etal}
\institute{
\inst{1} Physics Department, University of Fribourg, CH-1700 Fribourg, Switzerland\\
\inst{2} Web Sciences Center, School of Computer Science and Engineering,
University of Electronic Science and Technology of China, Chengdu 610054, P. R. China}

\abstract{Recommender systems recommend objects regardless of potential adverse effects of their overcrowding. We address this shortcoming by introducing crowd-avoiding recommendation where each object can be shared by only a limited number of users or where object utility diminishes with the number of users sharing it. We use real data to show that contrary to expectations, the introduction of these constraints enhances recommendation accuracy and diversity even in systems where overcrowding is not detrimental. The observed accuracy improvements are explained in terms of removing potential bias of the recommendation method. We finally propose a way to model artificial socio-economic systems with crowd avoidance and obtain first analytical results.}
\begin{document}

\maketitle

\section{Introduction}
Recommender systems are a powerful tool which nowadays helps most online retailers to make effective offers to their consumers. They use past user preferences to recommend new objects that the users might like. Research of recommendation grows rapidly and tackles issues like recommendation algorithms~\cite{AdoTuzi05,Lu12}, recommendation in social systems~\cite{Burke12}, and the use of recommendation in e-commerce~\cite{Schafer01}.

While there are situations where an arbitrary number of users can be recommended the same object, in other situations this is not the case. For example, one cannot recommend the same restaurant to many people as it has limited space and service capabilities. Similarly, it is not advantageous to use data on industrial production in countries~\cite{Pietronero01,Hidalgo01,Hidalgo02} and recommend the same new product to many countries as it could lead to undue competition and a poor ultimate outcome.

We propose crowd-avoiding recommendation which addresses this issue by imposing a strict occupancy constraint on individual objects or by assuming that object utility decays with the number of users sharing it. Our approach is linked to physics where particles occupy the energetically most favorable states but are either allowed in single occupation (fermions) or obey no restrictions (bosons). Although for particles there are no other options, we are here interested also in situations lying between these two extremes where an object can be recommended to an intermediate number of users. There is also a close connection with the combinatorial assignment problem where agents (users) can perform certain tasks (objects) with a certain cost and one searches for a bijective agent-task matching that minimizes the sum of the corresponding costs~\cite{Burkard09}. We use here some of the algorithms originally developed for the assignment problem.

Note that crowd avoidance is a general concept which can be used also
in situations where resources can be shared by an arbitrary number of
parties and user satisfaction does not decay with the number of other
users sharing the resource. Given a set of user score (or cost)
values, one can always apply an occupancy constraint or penalty and
see how this impacts the assignment of objects to users. It is of
particular importance to note that this new assignment is bound to be
more diverse than the original one where no additional constraints
were present. For example, if an individual object scores top for many
users, it cannot be assigned to all of them if the occupancy
constraint is sufficiently strong. Other objects then have to replace it
and the composition of the assigned objects becomes more diverse. In
this way, the crowd avoidance concept can help address one of the long
standing challenges of information filtering: the lack of
diversity~\cite{Ado12} and its potential adverse impact on network topologies~\cite{An01}.

In this letter, we study crowd avoidance and its effects in various socio-economic systems. We begin by defining possible approaches to introduce crowd avoidance in recommendation. We illustrate the use of this concept on empirical DVD renting data. Although one does not expect overcrowding to be a problem there, we show that including crowd avoidance in the recommendation process can increase the accuracy of the recommendation. To explain this unexpected observation, we use simple artificial data produced by a biased recommendation method and show that crowd-avoiding recommendation effectively removes this bias and thus increases recommendation accuracy. Finally we propose how to model artificial systems with crowd avoidance and suggest an analytical approach that can be used to study them.

\section{Framework}
We consider a set of $U$ users (which can be real persons but also firms or countries) and a set of $O$ objects (which can be restaurants, hotels, or sectors of industrial production). We then suppose that appreciation of object $\alpha$ by user $i$ is encoded in a single-valued utility $u_{i\alpha}$ (the higher the better). One can consider an idealized case where the true utility values are known or a case where some other information is used as a proxy for the utility. For example, recommendation scores obtained by a recommendation algorithm can serve this purpose---we shall study this in detail in the following section.

Our goal now is to model a system where for some reason it is not convenient for too many users to share the same object and thus some ``repulsion'' of users is in action. Reasons for this repulsion may vary and we shall discuss some of them in detail later on. Two approaches are at hand to model user repulsion. The first one (which we will refer to as ``effective utility'') postulates an effective user utility that decreases with the number of users $n_{\alpha}$ sharing object $\alpha$. One of the possible forms for the effective utility is
\begin{equation}
\label{continuous_utility}
\tilde u_{i\alpha}(n_{\alpha})=u_{i\alpha}/n_{\alpha}^b
\end{equation}
where the exponent $b$ determines the repulsion strength.\footnote{Other forms of effective utility, such as $\tilde u_{i\alpha}(n_{\alpha})=u_{i\alpha}-bn_{\alpha}$, are possible but we do not consider them here.}
When $b=0$, repulsion is absent and if all users prefer one single object, they are all free to herd on it. When $b\to\infty$, repulsion is extremely strong and utility-maximizing users prefer objects with the lowest occupancy over objects satisfying best their personal preferences. One can say that these two cases represent a bosonic and fermionic limit, respectively. The second approach (which we will refer to as 'constrained occupancy') postulates a rigid constraint $n_{\alpha}\leq m$ implying that each object can be shared by at most $m$ users. In terms of effective utility, this is equivalent to
\begin{equation}
\label{hardcore_utility}
\tilde u_{i\alpha}(n_{\alpha})=
\begin{cases}
u_{i\alpha} & \qquad n_\alpha\leq m,\\
0           & \qquad n_\alpha>m.
\end{cases}
\end{equation}

Given user preferences (represented by utility values $u_{i\alpha}$ and by how this utility diminishes with object occupancies $n_{\alpha}$), the natural next step is to find the best assignment of objects to users. Here the two important approaches are user-centered optimization where users attempt to maximize their own $\tilde u_{i\alpha}$ and global optimization where the sum of all effective utility values is maximized. User-centered optima can be obtained by a simple process where users arrive consecutively and choose their most preferred object (MPO) or by a more complicated process where users are allowed to change their choice until no one has an incentive to change and a Nash equilibrium is reached. These two processes have their real motivations: one might be tempted to leave a suddenly overcrowded bar for another one (corresponding to the Nash equilibrium case) but if one has already booked a good seat in a theater, there is no reason to care how many people did their bookings after (corresponding to the MPO case). Since multiple solutions can be found in all three optimization approaches (due to the degeneracy of both the global optimum and Nash equilibrium and due to the dependency of MPO on the users' order of arrival), all results presented here are averaged over several independent realizations.

Although constrained occupancy and effective utility have many features in common, they allow to view the problem from slightly different angles. The former approach has an analog in the classical \emph{assignment problem} where one wishes to find an optimal user-objects assignment with respect to a global energy function~\cite{Burkard09}. Fast algorithms exist to find global optima in this case and study their relation with user-centered optimization~\cite{Kuhn01}. Tools from spin glass theory are also of use here~\cite{Parisi85,Parisi86} (in our case, however, the true glassy phase is absent as the number of local minima grows only polynomially, not exponentially with system size). The effective utility approach instead allows, as we shall see later, for simpler analytical treatment.

\section{Evidence from empirical data}
We now demonstrate the concept of crowd avoidance on empirical DVD rental data released for the NetflixPrize (see \url{www.netflixprize.com}) from which we randomly choose $2,000$ users who rated at least $100$ objects in the original data and $2,000$ objects rated by at least $10,000$ users in the original data. The resulting set contains $592,995$ evaluations in the integer rating scale from $1$ to $5$. For the purpose of recommendation with unary data (\ie, without ratings), ratings $3$ or more are interpreted as favorable and constitute $515,342$ user-object links in the unary data set. Note that for the present case of users renting DVDs, the concept of crowd avoidance is relevant more for its ability to diversify the recommended content than for some real decline of utility when many users share a DVD (although a limited number of physical DVD copies might impact users by creating waiting times for popular objects).

The usual way to evaluate a recommender system is to move $10\%$ of
the data to a so-called probe set and then use the remaining $90\%$ of
the data (a so-called training set) to recover the missing (but known)
probe part~\cite{Lu12}. We use a variant of a popular and
high-performing recommendation method \emph{Singular Value
Decomposition} (SVD)~\cite{Takacs2007,Koren2009} for data with ratings and
the \emph{Probability Spreading} method (ProbS)~\cite{Zhou07,Zhou10} for
data without ratings. In both cases, the training set is used to obtain
estimated ratings or recommendation scores $u_{i\alpha}$ (the higher
the better) for all user-object pairs. Ultimately, a ranked list of
objects is created for each user $i$ where rank of object $\alpha$ is
$r_{i\alpha}$. In traditional recommendation, top $T$ objects from a
user's recommendation list are ``recommended''. One then evaluates
performance of the recommendations by comparing these top $T$ objects
with user-object pairs in the probe---the more of them appear among
the objects recommended to the given user, the better. The usual
corresponding metric is called \emph{precision} and it is defined as
the ratio of recommended probe objects to the total number of
recommended objects $TU$. Note that the standard procedure of creating the probe by random selection of links results in popular objects being over-represented in the probe.
This favors the ProbS method which is popularity-biased and disadvantages the SVD method. As a result, SVD under-performs ProbS despite the former using more information (\ie, the ratings) and thus being generally more reliable.
Since our focus here does not lie in a direct comparison of these two methods, this aspect is not essential for our analysis.

To evaluate the diversity of recommended objects, we use their effective number
\begin{equation}
\neff:=(TU)^2\Big/\sum_{\alpha=1}^O n_{\alpha}^2 
\end{equation}
where $n_{\alpha}$ is the number of users who get recommended object $\alpha$. When all objects are recommend equally often, $n_{\alpha}=TU/O$ and $\neff=O$; when the same object is recommended to all users, $\neff=1$.

To study the effect of crowd avoidance we turn $u_{i\alpha}$ into $\tilde u_{i\alpha}(n_{\alpha})$ and use the two above-described approaches (constrained occupancy and effective utility) to obtain a new set of recommendation lists. Since the number of evaluated top objects $T$ is found to be of little importance in our tests, we set $T=1$ for simplicity (thus, only one object is recommended to each user). The simplest approach to select the recommended objects is based on local optimization: users arrive in random order and choose their most preferred object (MPO), that is the one with the highest present effective utility. Since the outcome depends on the order of arrival of users, we always average our results over $1,000$ independent realizations of this process. It is straightforward to generalize this approach to a Nash equilibrium framework where user preferences are evaluated repeatedly and users are allowed to switch to another object which they prefer more than their originally selected object. Finally, it is also possible to consider global optimization where a global quantity, in our case the total effective utility, is maximized. Finding a globally optimal assignment of objects to users is a daunting task but fortunately, effective algorithms exist for the case of constrained occupancy. We use the classical Hungarian algorithm~\cite{Kuhn01} which, due to memory constraints, allows us to study the system for the occupancy constraint $m\leq 12$. Due to excessive computational complexity of the problem, we do not consider global optimization for $\tilde u_{i\alpha}$ given by Eq.~(\ref{continuous_utility}).

\begin{figure}
\centering
\includegraphics[scale=0.27]{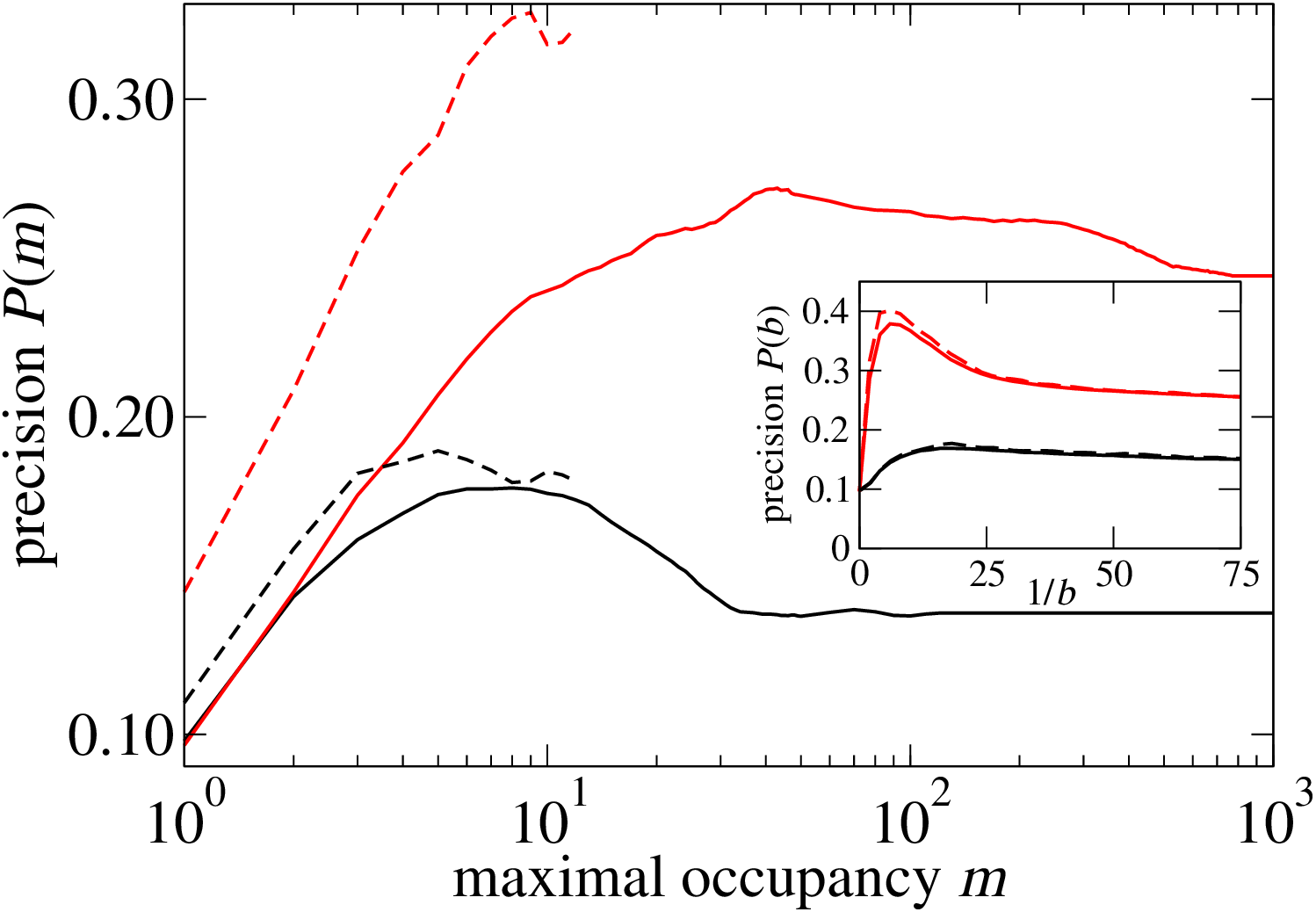}
\caption{Recommendation precision vs maximal occupancy $m$: black and red
lines denote results for data with ratings (SVD recommendation) and
data without ratings (ProbS recommendation), respectively. The solid
and dashed lines denote results for MPO and global optimization,
respectively. The inset shows precision vs $1/b$ with the solid and
dashed lines representing MPO and Nash equilibrium, respectively.}
\label{fig:real_data1}
\end{figure}

The resulting precision dependencies are shown in Fig.~\ref{fig:real_data1}. When $m\approx U$ or $b\approx 0$, the allowed occupancy is enough to accommodate all users or repulsion is weak and results are thus identical with the assignment of the object with the highest $u_{i\alpha}$ to each user. When $m=1$ or $b\to\infty$ (the fermionic limit), one is forced to assign much inferior objects to some users and the recommendation precision suffers. However, the course of precision is not monotonous: when some intermediate occupancy constraint is applied, precision can be improved and this improvement is further magnified if a sophisticated optimization scheme (Nash/global) is used.

\begin{figure}
\centering
\includegraphics[scale=0.3]{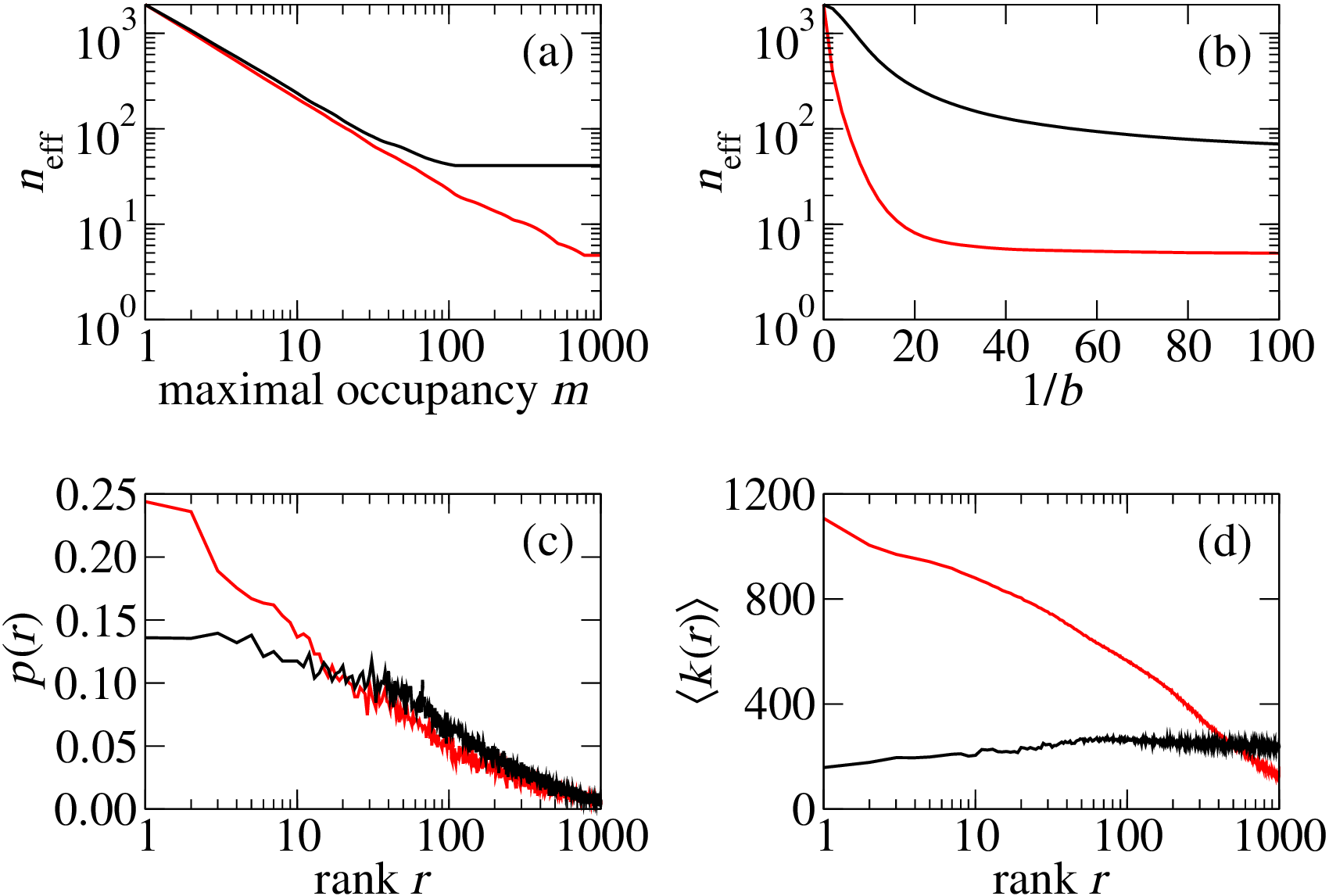}
\caption{Panels (a) and (b) show the effective number of recommended
objects vs maximal occupancy and $1/b$, respectively. Panels (c) and
(d) show the average recommendation precision and the average object
degree vs object rank in the recommendation lists, respectively. Red
lines refer to results obtained by ProbS and black lines refer to
those obtained by SVD.}
\label{fig:real_data2}
\end{figure}

Fig.~\ref{fig:real_data2}a,b show the effective number of
objects recommended to users, $\neff$,  which grows with repulsion
strength as expected. At $m=14$ which maximizes the
precision value $P(m)$ for SVD recommendation, the observed
$\neff\approx155$ is significantly higher than $\neff\approx11$
achieved when the occupancy constraint is missing. We can thus
conclude that the artificial occupancy constraint allows us to
simultaneously improve precision and diversity of recommendation
lists. Furthermore, one can easily show that the observed precision improvement
is not just an artifact of using recommendation methods which do not
put best objects to the top of their recommendation lists.
Fig.~\ref{fig:real_data2}c shows precision $P(r)$ when objects ranked
$r$ are recommended to each user without any regards to occupancy and
repulsion and demonstrates that the further down the recommendation list
we go, the lower the achieved precision. Why then crowd avoidance
helps?

When sufficiently strong occupancy restriction is applied, one is
forced to move in recommendation lists of some users from the top rank $1$ to a
lower rank $r$. This move does not influence the precision if the
top-ranked object and the newly selected object are both probe or both
non-probe. Therefore to observe a precision improvement, the
probability of exchanging a non-probe object at rank $1$ for a probe
object at rank $r$ must be greater than the probability of exchanging
a probe object at rank $1$ for a non-probe object at rank $r$. Such a
situation can occur if the used recommendation method is biased in
some way and, along with successful recommendations demonstrated by
Fig.~\ref{fig:real_data2}c, places some wrong objects at the top of
many users' recommendation lists. For example,
Fig.~\ref{fig:real_data2}d shows that ProbS is strongly biased towards
popular objects that tend to end up at the top of recommendation lists
(at the same time, SVD is not popularity biased or it is even weakly
biased in the opposite direction).

The question still remains of how such advantageous channeling and recommendation bias arise. To approach it, we construct a set of artificial recommendation lists producing the same phenomenon. Each artificial object is assigned a random hidden variable $h_{\alpha}$ which encodes a particular characteristic of the object (popularity or something less tangible) to which a recommendation method can be sensitive and biased. For example, the general bias of recommender systems towards popular objects is well known and represents one of open problems in this field~\cite{Celma08,Zhou10}.
A recommendation list for a given user contains a small number of probe objects which are chosen from all objects at random and the rest are non-probe objects. The ranking of all objects is then constructed from top to bottom by applying two simple rules: (i) With probability $Q(r)$, a probe object is chosen at random from the remaining probe objects and placed at rank $r$; Here $Q(r)$ is a monotonically decreasing function of rank $r$; (ii) In the opposite case, a non-probe object is chosen from the remaining non-probe objects with probability proportional to $h_{\alpha}$.
In effect, the first rule states that the recommendation method used to build the lists works well and puts probe objects preferentially at the top of recommendation lists. The second rule implies that errors of the recommendation method are biased by the hidden variable $h_\alpha$. To check whether the second assumption is necessary, we also present results for the case where non-probe objects are chosen purely at random.

\begin{figure}
\centering
\includegraphics[scale=0.27]{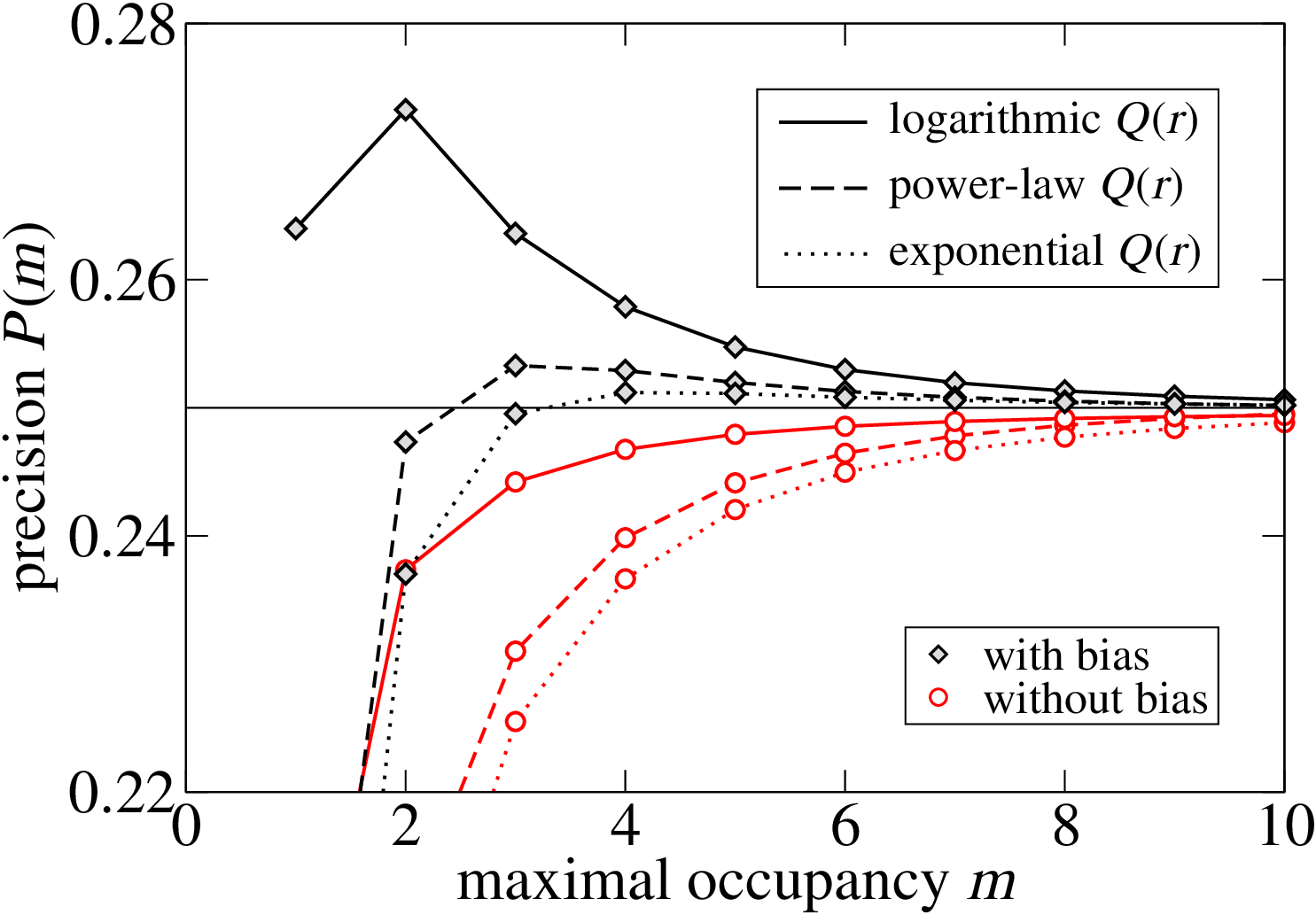}
\caption{Results as in Fig.~\ref{fig:real_data1} for the artificial data described in the text with and without the popularity bias. The probe object occurrence probability decay as: $Q(r)=p(1)[1-\ln r/\ln r_{\mathrm{max}}]$ (solid lines, $Q(r)=0$ for $r\geq r_{\mathrm{max}}$), $Q(r)=[p(1)-\epsilon_1]/r+\epsilon_1$ (dashed lines), and $Q(r)=[p(1)-\epsilon_2]\,\mathrm{e}^{1-r}+\epsilon_2$ (dotted lines) where $p(1)=0.25$. Parameters $r_{\mathrm{max}}$, $\epsilon_1$, and $\epsilon_2$ are set in order to have an average of $10$ probe objects for each user, \ie, $\sum_r Q(r)=10$. With bias, $P(m)$ shows a maximum at $m<U$ in all three cases.}
\label{fig:artificial}
\end{figure}

Artificial datasets were created for $2,000$ users and $2,000$ objects
as in our real data. $h_{\alpha}$ values were drawn from the standard
log-normal distribution with mean $0$ and sigma $1$. We then run MPO
for artificial datasets and average over multiple realizations of
recommendation lists and the order of arrival of users.
Fig.~\ref{fig:artificial} shows that when bias is present, constrained
occupancy can enhance the recommendation precision. The shape and
magnitude of this enhancement depends on $Q(r)$ (note that in
particular the logarithmically-decaying $Q(r)$ is supported by how
recommendation precision decays when only objects of a particular ranking
$r$ are chosen---this is shown as $p(r)$ in Fig.~\ref{fig:real_data2}c)
and on the distribution of $h_{\alpha}$ (the narrower the distribution,
the weaker the bias and the smaller the possible gain due to constrained
occupancy). When the bias is removed entirely, no precision
improvements can be seen and $P(m)$ grows monotonically with $m$ (\ie,
the weaker the occupancy constraint, the higher the precision).
Similarly, if the choice of both probe and non-probe objects is
subject to the same bias, precision improvement vanishes too (the two
groups of objects then essentially merge). If probe and non-probe
objects are subject to different kinds of bias, precision improvement
can be still achieved.

To summarize, we found that if a recommendation algorithm has some
bias, then introducing user repulsion or constrained occupancy not
only helps to increase the diversity of recommendations but it may
also enhance their precision. This enhancement can be substantial if
the actual bias disagrees with preferences of the users but it can
also vanish if the two are in line (\eg, if an algorithm is biased
towards popular objects and the users \emph{want} popular objects).
Note that we studied data where no real preference for low object
occupancy is expected. In the opposite case, one could expect even stronger positive effects
from introducing crowd avoidance.

One can generalize Eq.~(\ref{hardcore_utility}) to a heterogeneous occupancy
constraint by imposing $n_{\alpha}\leq m_{\alpha}$. For example, one can set
$m_{\alpha}=C k_{\alpha}^l$ where $k_{\alpha}$ is popularity (degree) of object $\alpha$
in the input data, exponent $l$ determines the level of heterogeneity,
and $C$ allows for fixing the average maximum occupancy at a given value
$m$. $l=0$ corresponds to the already discussed case where all $m_{\alpha}=m$.
Preliminary results for the SVD method show that with $l>0$ it is
possible to further magnify the precision improvement to a level similar to that of ProbS
(that is, precision values around $0.25$). This is in line with the previous observation that SVD suffers of the way the probe set is selected. If popular objects can be shared
by more users than unpopular ones (\ie, when $l>0$), they are more likely to be recommended
and thus the precision can be further improved. For the ProbS method, we observe no substantial precision improvement for $l\neq0$.

\section{Analysis of artificial data}
In addition to the described practical applications, a complementary view can be obtained by studying artificial systems where crowd avoidance plays a role. To do that, we replace empirical user preferences with artificial correlated utility values
\begin{equation}
u_{i\alpha}=\sqrt{1-c}\,x_{i\alpha}+\sqrt{c}\,q_{\alpha}
\end{equation}
where $q_{\alpha}$ represents the intrinsic quality of object $\alpha$, $x_{i\alpha}$ quantifies the individual preferences of user $i$ for object $\alpha$, and $c\in[0,1]$ determines the magnitude of user-user correlations. Elements $x_{i\alpha}$ and $q_{\alpha}$ are drawn independently from the standard normal distribution $\Nd(0,1)$ and thus the distribution of $u_{i\alpha}$ is also $\Nd(0,1)$ independently of the value of $c$. The Pearson correlation of scores given to an object by two different users is $C(i,j)=c$. We focus here on the case represented by Eq.~(\eqref{continuous_utility}) where the effective utility gradually decreases with the number of users sharing an object. Note that the term $n_\alpha^{-b}$ in $\tilde u_{i\alpha}$ affects only the magnitude not the sign of the effective utility. Objects that have negative score $u_{i\alpha}$ for a given user are therefore bound not to be chosen by this user regardless of their occupancy and occupancy of the other objects.

\begin{figure}
\centering
\includegraphics[scale=0.27]{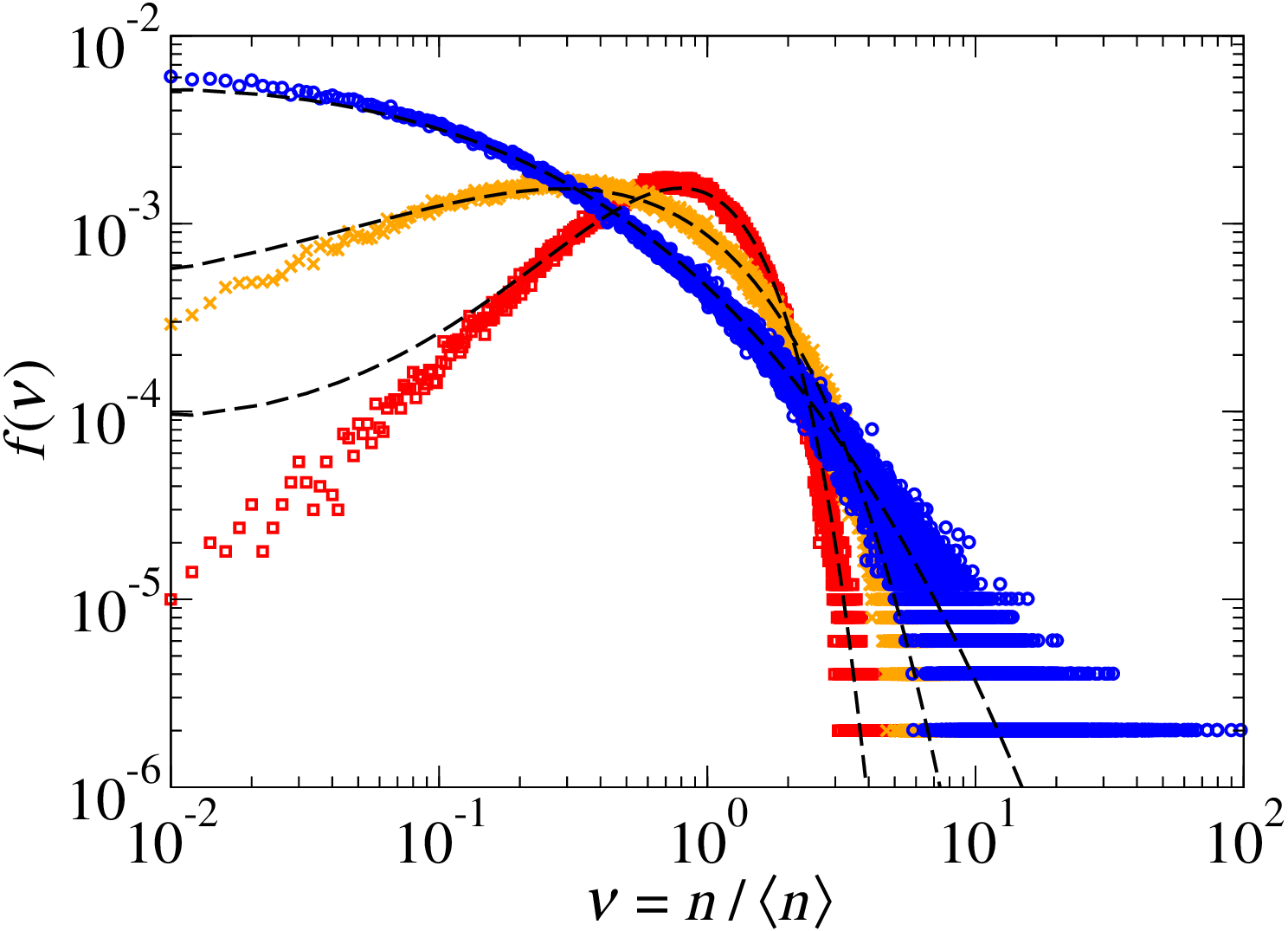}
\caption{Degree distribution for three values of the repulsion parameter $b$: $b=0.5$ (red squares), $b=0.25$ (orange crosses). $b=0.1$ (blue circles). Dashed black lines correspond to Eq.~(\ref{f_n}) (values of $F$ are respectively $2.5$, $3.0$ and $4.5$). The simulation parameters are $c=0.5$, $U=5\cdot10^5$ and $O=10^3$; $\avg{n}=U/O$.}
\label{fig:f_n}
\end{figure}

The first step is to compute the resulting object degree distribution $f(n_\alpha)$ for different levels of correlation and values of the repulsion strength. The relation between object utility $u_{i\alpha}$ and degree $n_{\alpha}$ is particularly simple when user correlation is perfect and thus $u_{i\alpha}=q_{\alpha}$. In the case of user-centered optimization (MPO/Nash), users gradually arrive and select the object with the currently highest effective utility. This flattens the effective utility landscape and the effective utility $q_{\alpha}/n_{\alpha}^b$ is thus constant across all objects in the limit of a large number of users, implying $q_\alpha=A(n_\alpha/\avg{n})^b$ where $A$ is determined by requiring $\sum_{\alpha} n_{\alpha}=U$ (each user selects one object). When user preferences are not perfectly correlated, a similar process takes place but the situation is more complicated. When a user selects an object, not only the intrinsic quality $q_{\alpha}$ but also the individual user preferences $x_{i\alpha}$ matter. One can illustrate their effect on an example of an object with $q_{\alpha}=0$ which is obviously chosen by no user when $c=1$. When $c<1$, positive $x_{i\alpha}$ terms make effective utility of this object positive for some users and $n_{\alpha}>0$. This leads us to a generalized dependency between $q_{\alpha}$ and $n_{\alpha}$ which has the form $q_{\alpha}+F=A(n_\alpha/\avg{n})^b$ where $F$ reflects fluctuations of $u_{i\alpha}$ around $q_{\alpha}$ and $A$ again makes it possible to achieve $\sum_{\alpha} n_{\alpha}=U$. Since the distribution of $q_{\alpha}$ is known, this relation directly leads to the degree distribution
\begin{equation}
\label{f_n}
f(\nu)=\frac{bA\nu^{b-1}}{\sqrt{2\pi(1-c)}}\exp\Big(-\frac{(A\nu^b-F)^2}{2(1-c)}\Big)
\end{equation}
where $\nu:=n/\avg{n}=nU/O$. One can approximate $F$ by extreme statistics for a normally distributed variable~\cite{Lu09} or by relating it to the fluctuation magnitude occurring with the probability $1/U$ (thus for one user on average). Since these approximations are too crude to produce good fits of the data, we choose $F$ by hand instead and then adjust $A$ accordingly. Fig.~\ref{fig:f_n} shows good agreement between this semi-analytical result and degree distributions following from numerical MPO matching in a system with $1,000$ objects, $500,000$ users, and $c=0.5$.

\section{Discussion}
Although recommending the same object to too many users can have adverse
effects in many real situations, this letter is to our best knowledge
the first attempt to introduce and study crowd-avoidance in recommender
systems. We showed that applying object occupancy
constraints or user repulsion to results of an ordinary recommendation
method can be beneficial in two complementary ways. Firstly, it
naturally increases the diversity of the recommended content and thus
helps to address one of the long standing issues in information
filtering~\cite{Zhou10}. Recommendation diversity is further closely
connected to sustainability of information filtering tools which emerges
as an important challenge~\cite{An01}. Secondly, crowd avoidance can
improve accuracy of the resulting recommendations---despite the fact that
herding of users does not reduces their benefits in the studied DVD
rental data. The practical problem of choosing a well-performing
constraint can be solved by parameter fine-tuning on a given data (by
hiding 10\% of the data in the same way as we did here) or by applying
some simple rules of thumb which are yet to be found.

It is rare that introducing constraints to an optimization problem can improve quality of the solution. We proposed a simple explanation for the unexpected accuracy improvement observed in our case which is based on correcting biases (whatever they are) of the recommendation algorithm. In some sense, crowd avoidance could serve to quantify the bias of a recommendation algorithm (the bigger the improvement from using crowd avoidance, the more biased the algorithm). The accuracy improvements disappear if the bias of the recommendation algorithm is too weak or if it is coupled with true users preferences.  One can for example choose a faction $1-\lambda$ of probe objects according to the bias given by the hidden variables $\{h_\alpha \}$ and the remaining $\lambda$ fraction of probe objects at random. In this case, there is a limit value $\lambda^*$ under which the accuracy improvement disappears. To derive simple analytical conditions under which the diversification would become disadvantageous remains a challenge for future research.

In systems where herding of users on an object reduces their benefits,
crowd-avoidance can be applied to find a good compromise between
satisfying the preferences of users and distributing them among objects
evenly. We presently lack real data where some utility decline with
occupancy can be expected. We thus studied artificially generated data
and found an approximate solution for the object degree distribution as
a function of the repulsion parameter.
The mathematical formalization of a crow-avoiding recommendation establishes
close connections with different areas, such as optimization
algorithms, spin glasses, and game theory, which could all contribute
to future progress and insights.

From the theoretical point of view, one can consider the case where user repulsion is replaced by attraction, possibly leading to a condensation phenomenon where a majority of users choose one object only (a similar situation has been found in the preferential attachment model with heterogeneous fitness values~\cite{BiBa01}). The situation becomes more interesting when the deterministic utility maximization is replaced by probabilistic choice. (The most straightforward way to do that is to assume probability of choosing an object with utility $u_{i\alpha}$ to be proportional to $\exp[\beta u_{i\alpha}$].) Various forms of user attraction can then manifest themselves in positive feedback mechanisms such as preferential attachment and its variants~\cite[Ch.~14]{Newman2010}. This effectively extends the fermion-boson range discussed here into a more complete one: fermion-boson-preferential attachment.

\acknowledgments
This work was supported by the EU FET-Open project QLectives (grant no.\ 231200) and by the Swiss National Science Foundation (grant no.\ 200020-132253). We acknowledge useful comments from Chi Ho Yeung, Joseph Wakeling, and an anonymous reviewer.

\end{document}